\documentclass[runningheads]{llncs}
\usepackage{fancyhdr}
\fancypagestyle{firstpage}{
  \fancyhf{}
  \fancyfoot[L]{
    \footnotesize
    This preprint has not undergone peer review or any post-submission improvements or corrections.
    The Version of Record of this contribution is published in Euro-Par 2026: Parallel Processing, and is available online at \url{https://doi.org/10.1007/978-3-032-35251-4_40}
  }
}
\makeatletter
\def\sectionmark#1{}%
\def\subsectionmark#1{}%
\makeatother

\usepackage{hyperref}
\hypersetup{
  colorlinks=true,
  allcolors=blue,
  bookmarksdepth=2
}

\usepackage{amsmath}
\usepackage{csquotes}
\usepackage{booktabs}
\usepackage{subcaption}
\usepackage{tikz}
\usetikzlibrary{graphs}
\usetikzlibrary{graphs.standard}
\usetikzlibrary{matrix}
\usetikzlibrary{fit}
\usetikzlibrary{external}
\usepackage{pgfplots}
\pgfplotsset{compat=1.18}
\usepackage{pgfplotstable}
\usepgfplotslibrary{groupplots}

\newif\iffigsprebuilt
\figsprebuilttrue
\iffigsprebuilt\else
\fi
\usepgfplotslibrary{colorbrewer}
\pgfkeys{
  /datafile/.is family, /datafile,
  default/.style = {
    prefix = plots/indirection/indirection,
    graph = 0,
    config = 0
  },
  prefix/.estore in = \dfprefix,
  graph/.estore in = \dfgraph,
  config/.estore in = \dfconfig
}

\newcommand{\datafileKV}[1]{%
  \pgfkeys{/datafile, default, #1}%
  \edef\kascadedatafilename{\dfprefix-data-in\dfgraph-c\dfconfig.csv}%
}
\pgfplotscreateplotcyclelist{lineplotlist}{
  {Dark2-A,mark=*,mark options={scale=0.8}},
  {Dark2-B,mark=square*,mark options={scale=0.8}},
  {Dark2-C,mark=triangle*},
  {Dark2-D,mark=diamond*},
}
\pgfplotsset{
  cycle list name=lineplotlist,
  kascade plot format/.style={
    /pgfplots/table/x=p_exp,
    /pgfplots/table/y=total_time_mean,
    /pgfplots/table/y error plus expr=\thisrow{total_time_max}-\thisrow{total_time_mean},
    /pgfplots/table/y error minus expr=\thisrow{total_time_mean}-\thisrow{total_time_min},
    /pgfplots/table/col sep=comma,
  }
}

\newcommand{\Tallall}{T_{\mathrm{all2all}}}
\newcommand{\Tdoubling}{T_{\mathrm{doubling}}}

\newcommand{\Tbase}{T_{\mathrm{base}}}
\newcommand{\Oh}[1]{\mathcal{O}\!\left( #1\right)}
\newcommand{\Ohsmall}[1]{\mathcal{O}(#1)}
\newcommand{\Th}[1]{\mathrm{\Theta}\!\left( #1\right)}
\newcommand{\Thsmall}[1]{\mathrm{\Theta}(#1)}
\newcommand{\Seq}[2]{#1..#2}
\newcommand{\SetOfLists}{\mathcal{L}}
\spnewtheorem{observation}[theorem]{Observation}{\bfseries}{\itshape}

\usepackage[linesnumbered,vlined,ruled]{algorithm2e}
\colorlet{my-hl1}{Set1-B}
\colorlet{my-hl2}{Set1-D}

\SetCommentSty{mylistingcommentstyle}
\DontPrintSemicolon
\SetKwFor{OnMessage}{on message}{do}{end}
\SetKwFor{Request}{request}{then}{end}
\SetKw{SendKw}{send}
\SetKw{ToKw}{to}
\SetKw{FromKw}{from}
\newcommand{\Send}[2]{\SendKw #1 \ToKw #2}
\SetKwArray{succ}{succ}
\SetKwArray{rank}{rank}
\SetKwFunction{owner}{owner}

\usepackage{microtype}
\usepackage[capitalise]{cleveref}
\makeatletter
\AtBeginDocument{%
  \let\kascade@oldlabel\cref@old@label
  \def\cref@old@label#1{%
    \let\kascade@creflabel\cref@currentlabel
    \kascade@oldlabel{#1}%
    \let\cref@currentlabel\kascade@creflabel
  }%
}
\makeatother

\crefname{line}{line}{lines}
\Crefname{line}{Line}{Lines}
\crefname{observation}{Observation}{Observations}

\usepackage{orcidlink}
\newcommand{\myorcid}[1]{\orcidlink{#1}}
\usepackage[misc]{ifsym}
\newcommand{\corres}[1]{
  \textsuperscript{(\href{mailto:#1}{\color{black}\Letter})}
}

\usepackage[T1]{fontenc}
\usepackage{graphicx}
\usepackage{color}
\usepackage[textsize=scriptsize]{todonotes}
\todostyle{nu}{color=red,backgroundcolor=red!20,bordercolor=red,author=NU}
\todostyle{ms}{color=blue,backgroundcolor=blue!20,bordercolor=blue,author=MS}
\todostyle{ps}{color=green,backgroundcolor=green!20,bordercolor=green,author=PS}
\begin{document}
\title{Engineering Scalable Distributed List Ranking}
\author{
  Peter Sanders \myorcid{0000-0003-3330-9349} \and
  Matthias Schimek \myorcid{0009-0002-6402-9016} \and\\
  Tim Niklas Uhl \corres{uhl@kit.edu}\myorcid{0000-0001-9295-1388} \and
  Thomas Weidmann
}
\authorrunning{Sanders, Schimek, Uhl, Weidmann}
\institute{%
  Karlsruhe Institute of Technology, Karlsruhe, Germany\\
  \email{\{sanders,schimek,uhl\}@kit.edu}\\
  \email{thomas.weidmann@outlook.com}
}
\pagestyle{fancy}
\maketitle              %
\thispagestyle{firstpage}
\setcounter{footnote}{0}
\begin{abstract}
  The list ranking problem is one of the classical problems of parallel computing, with nontrivial algorithms and many applications as a subroutine for solving other problems.
  While it has been intensively studied in the early days of parallel computing, few things happened in the last 20 years.
  In particular, there is little work on scaling list ranking to large machines and input sizes.
  We reconsider list ranking starting from the ground-breaking results of Sibeyn a quarter century ago.
  We employ algorithm and performance engineering to improve his sparse ruling-set algorithm, making it capable of scaling to many processors, and provide a more detailed analysis of the impact of the algorithm's parameters, further guiding our practical implementation.

  We perform an extensive experimental study across a variety of input instances with different structural properties.
  We demonstrate that indirect communication, exploiting input locality, and message coalescing allows scaling to billions of elements on up to 24\,576 cores.

\end{abstract}

\section{Introduction}\label{sec:introduction}
List ranking is one of the most fundamental tasks in parallel processing, where we ask for
the distance of list elements to the end of their list.
This can be used to compute prefix sums on lists, converting lists to arrays, or to perform basic operations on trees (and graphs) using
the Euler tour technique \cite{Jaja1992}. More concrete applications can be found in bioinformatics, e.g.\ for chain compaction in de Bruijn graph based genome assembly~\cite{Pan2020}.

There is a trivial sequential linear-time algorithm.
\emph{Pointer chasing} basically just traverses the lists.
However, designing fast and work-optimal parallel algorithms is challenging.
Historically, the list ranking problem has been extensively studied in parallel computing, starting with Wyllie's simple, fast, yet suboptimal \emph{pointer doubling}  \cite{wyllie1979complexity}.
Subsequent work achieved work-optimality and logarithmic time on PRAMs \cite{miller1985parallel,DBLP:conf/focs/ColeV86}.
Most of these algorithms share the same general approach: They splice out elements of the initial list, solve the remaining smaller subproblem, and then reinsert the removed elements, updating their ranks.
For example, \emph{independent set removal} (ISR) splices out a constant fraction of nonadjacent elements \cite{DBLP:journals/ipl/AndersonM90}. However, many of these algorithms are not work efficient with respect to the involved constant factors and thus not a good basis for a practical implementation.

A more efficient alternative is to remove all but a sparse set of \emph{ruler} elements \cite{reid1994list,anderson1991deterministic}. The sublists between two rulers are spliced out using pointer chasing. Thus, most work in this \emph{sparse ruling set} (SRS) algorithm is as efficient as the sequential algorithm. 
This makes SRS attractive as the basis for designing efficient distributed memory algorithms.
The arguably most comprehensive line of work in this research direction is due to Sibeyn~\cite{sibeyn_list-ranking_2003,Sibeyn1999a,sibeyn_better_1997,DBLP:conf/hipc/Sibeyn99,sibeyn2003minimizing}, presenting distributed-memory adaptations of SRS, ISR, and pointer doubling and a practical evaluation on random input lists on up to 130 processors~\cite{Sibeyn1999a}.
This evaluation shows that the SRS algorithm is the most efficient one for sufficiently large input sizes.

Our work revisits the most efficient distributed-memory algorithm, SRS with spawning~\cite{DBLP:conf/hipc/Sibeyn99}, providing the first practical evaluation of it.
We show that it scales across a large variety of inputs by considering input locality, machine topology, and reducing the number of communication start-ups and the number of communication rounds.

\paragraph{Our Contributions.}
\begin{itemize}
\item We provide the first practical systematic evaluation of sparse ruling-set with spawning.
\item We scale to two orders of magnitude more processors than previous studies on general purpose list ranking, demonstrating scalability up to 24\,576 cores.
\item We use a variety of input instances with and without locality.
\item We show that hardware-topology aware indirection, exploiting locality, and message coalescing are essential to achieve scalability.
\item We provide a refined scalability analysis of sparse ruling-set with spawning which considers messages indirection and locality. This helps us to choose good parameters in practice.
\end{itemize}

\subsection{Problem Definition and Notation}\label{sec:preliminaries}
The \emph{list ranking problem} is defined as follows:
As input we consider a set $\SetOfLists$ of singly-linked lists, with elements labeled $\Seq{0}{n-1}$\footnote{$\Seq{a}{b}$ is shorthand for the sequence $[a, a+1, \ldots, b-1, b]$} . We call a list element \emph{initial} if it has no predecessor, and \emph{terminal} if it has no successor.
The list ranking problem now asks for identifying each list element $i$ with the terminal element $i_{\text{term}}$ of its list, and to determine the number of links to follow to reach $i_{\text{term}}$ from $i$.
This is denoted as the \emph{rank} of $i$.

More concretely, $\SetOfLists$ is represented as a \emph{successor array} \succ{$\Seq{0}{n-1}$} of size $n$, where \succ{$i$} denotes the successor of element $i$.
Terminal elements are indicated by pointing to themselves, i.e., $\succ{$i$}=i$.
A list ranking algorithm will then modify \succ, such that \succ{$i$} points to the terminal element of $i$.
The element ranks are stored in an array \rank{$\Seq{0}{n-1}$}.

A generalization of this is the \emph{weighted list ranking problem} where each link to a successor \succ{i} is additionally associated with a weight.
Then \rank{$i$} is defined as the weighted distance from $i$ to its terminal element, i.e., the sum of the weights over the path from $i$ to the terminal element.
We assume that the weight of $i$ is initially stored in \rank{$i$}, and the array is modified in-place.
Note that an instance with input $\rank{$i$} = 0$ for terminal elements and $\rank{$i$} = 1$ for all other elements is equivalent to the unweighted case.

\subsection{Machine Model and Input Format}\label{sec:machine-model-input}
We consider a distributed-memory system consisting of $p$ processing elements (PEs) numbered $\Seq{0}{p-1}$ allowing single-ported point-to-point communication between arbitrary communication partners~\cite{SMDD19}.
Sending a message of length $\ell$ from one PE to another takes time $\alpha + \beta\ell$, where $\alpha$ is the time required to initiate a connection and $\beta$ the subsequent transmission time for sending a machine word.

We assume that the input and output arrays are stored equally distributed across the PEs, such that PE $j$ is responsible for a contiguous index range $\Seq{a_j}{b_j}$ of size $\Thsmall{n/p}$ and holds the partial data \succ{$\Seq{a_j}{b_j}$}, \rank{$\Seq{a_j}{b_j}$}.
We call $\Seq{a_j}{b_j}$ the \emph{local} elements of PE $j$ and define $\owner{$i$} = j$ if element $i$ is local to PE $j$.
All other elements are \emph{remote} elements for PE $j$.
PE $j$ can only access local elements directly, for remote elements it explicitly sends a request.

\subsection{Other Related Work}\label{sec:other-related-work}
Sparse ruling-set and independent set removal have also been adapted for the coarse-grained bulk-synchronous CGM and BSP computational model~\cite{dehne1997randomized,dehne2002efficient,DBLP:journals/jea/LassousG02} and evaluated in practice.
The problem has also been examined on GPUs~\cite{fast_and_scalable_gpu,wei2012optimization}.%

Apart from bulk-synchronous algorithms, Sibeyn proposed an approach using $p$ communication rounds, where each processor communicates with at most one or two others per round.
This results in expected communication volume in $\Ohsmall{n/p \ln{p}}$ and an overall number of $\Ohsmall{p}$ messages sent and received by each processor for random inputs~\cite{sibeyn_better_1997}, outperforming plain pointer-doubling in practice.
A related idea with similar bounds has been described by Träff~\cite{traff_portable_1998}.

\section{Engineering a Scalable List Ranking Algorithm}\label{sec:main-contribution}
We will first outline the sparse ruling-set algorithm in detail, and then focus on how to adapt this approach to scale to today's largest machines by exploiting locality (\cref{sec:locality}), employing message indirection (\cref{sec:indirection}), and additional practical consideration (\cref{sec:further-optimizations}).
Many of these optimizations can also be applied to pointer doubling and independent set removal to improve scalability, but the algorithmic properties of sparse ruling-set make it a prime candidate for ranking lists at scale.
Finally, we analyze the scalability of our algorithm and derive an optimal ruler selection strategy in \cref{sec:ruler-selection}.

\subsection{The Sparse Ruling-Set Algorithm}\label{sec:sparse-ruling-set}
As introduced in \cref{sec:introduction} sparse ruling-set aims to reduce the problem size by running multiple sequential list-traversals in parallel from a set of starting elements (called \emph{rulers}).
When such a search reaches another ruler, we compress the list segment between the two rulers to a single link with the total weight of the segment. Then, we recurse on the subproblem consisting only of rulers.
Once the subproblem has been fully ranked, we reinsert the compressed list segments between rulers and adjust ranks and terminal elements accordingly.

\begin{figure}[t]
  \begin{algorithm}[H]
  \SetKwFunction{PickRulers}{pickRulers}
  \SetKwFunction{FindLeaves}{findInit}
  \SetKwFunction{ListRanking}{listRanking}
  \SetKwFunction{ReverseList}{reverseList}
  \SetKwFunction{Owner}{owner}

  \KwData{\succ{$\Seq{a_j}{b_j}$}, \rank{$\Seq{a_j}{b_j}$}}
  \KwIn{PE index $j$, number of desired rulers $r$}
  $(\succ,\rank) \leftarrow \ReverseList{\succ, \rank}$\;
  $I_j \leftarrow \FindLeaves{\succ}$;
  $r_j \leftarrow r\frac{b_j - a_j}{n}$;
  $R_j \leftarrow \PickRulers{$r_j-|I_j|, \Seq{a_j}{b_j}$} \cup I_j$\;\label{fig:srs-algo:line:ruler-selection}
  
  \ForEach(\tcp*[h]{Ruler chasing}){$r \in R_j$}{
    \Send{$(\rank{r}, \succ{r}, r)$}{\Owner{\succ{$r$}}}\;
    $\succ{$r$} \leftarrow r$;
    $\rank{$r$} \leftarrow 0$\;
  }
  \While{messages to send}{\label{fig:srs-algo:line:ruler-loop-begin}
    \OnMessage{$(w, i, r)$}{

      \uIf(\tcp*[h]{Ruler-spawning}){$i$ is terminal or ruler}{
        select unvisited node $r_0 \in \Seq{a_i}{b_i}$ as new ruler\;\label{fig:srs-algo:line:ruler-spawning-begin}
        $\succ{$r_0$} \leftarrow r_0$;
        $\rank{$r_0$} \leftarrow 0$;
        $R_j \leftarrow R_j \cup \{r_0\}$\;
        \Send{$(\rank{$r_0$}, \succ{$r_0$}, r_0)$}{\Owner{\succ{$r$}}}\;\label{fig:srs-algo:line:ruler-spawning-end}
      }
      \Else{
        \Send{$(w + \rank{i}, \succ{i}, r)$}{\Owner{\succ{$i$}}}\;
      }
      $\succ{i} \leftarrow r$;
      $\rank{i} \leftarrow w$\;
    }} \label{fig:srs-algo:line:ruler-loop-end}

  \ListRanking{\succ{$R_j$}, \rank{$R_j$}}\tcp*[l]{Base case}\label{fig:srs-algo:line:base-case}  
  \ForAll(\tcp*[h]{Ruler propagation}){$i \in \Seq{a_j}{b_j}$ that are no rulers}{
    \Request{$(\succ{\succ{i}}, \rank{\succ{i}})$ \FromKw \Owner{\succ{i}}}{
      $\succ{i} \leftarrow \succ{\succ{i}}$\;
      $\rank{i} \leftarrow \rank{\succ{i}} + \rank{i}$\;
    }
  }
  \caption{The sparse ruling-set algorithm.}
  \label{fig:srs-algo}
\end{algorithm}
\end{figure}

\cref{fig:srs-algo} describes sparse ruling-set in more detail:
First, we randomly sample a set of $r$ rulers.
This can be done without communication by randomly sampling $r/p$ elements per PE\@.
Note that we also need to include all initial elements in the set of rulers to guarantee that the full list is traversed (\cref{fig:srs-algo:line:ruler-selection}).
In the subsequent \emph{ruler-chasing phase}, we start packet waves from the rulers:
For each ruler $r$, we send a message consisting of ($\rank{$r$}$, $\succ{$r$}$, $r$) to the PE owning $\succ{$r$}$ and set $\rank{$r$} = 0$ and $\succ{$r$} = r$.
Upon receiving a message $(w, i, r)$ dedicated to element $i$, we update the \succ and \rank value of $i$ accordingly with $r$ and $w$ and forward the message to $\succ{$i$}$, increasing the transmitted weight by \rank{$i$}.
When $i$ is a ruler or a terminal element, we still update the values, but do not send a new packet.
A straightforward way of implementing the communication during ruler chasing (\crefrange{fig:srs-algo:line:ruler-loop-begin}{fig:srs-algo:line:ruler-loop-end}) is to use bulk-synchronous communication rounds. %
However, this has a major shortcoming: The number of rounds is determined by the maximum length of a sublist between two rulers.
In case of bad ruler selection, this may lead to load imbalances which may hinder scalability.
This was addressed by Sibeyn~\cite{DBLP:conf/hipc/Sibeyn99} using so-called \emph{ruler-spawning}, which we will discuss in \cref{sec:ruler-spawning}.

After ruler chasing, each element points to the ruler it was reached from, including rulers themselves. %
We interpret the \succ array, limited to rulers, as a new weighted list ranking problem and apply a \emph{base-case} list ranking algorithm to compute the distance from each ruler to the initial element of its list.

Finally, we need to fix the rank and initial element of the elements that have been eliminated by ruler chasing.
In the \emph{ruler-propagation} phase, each non-ruler element $i$ sends a request to the ruler it was reached from, which is stored in \succ{$i$}.
A ruler $r$ then replies with its initial node \succ{$r$} distance \rank{$r$}, and each non-ruler updates its local result accordingly. %
Since ruler chasing effectively reverses the list direction in the subproblem, the algorithm as described above computes ranks with respect to the initial instead of terminal elements.
This can be fixed by reversing the list as a preprocessing step, effectively using the predecessor array as input.

\subsection{Ruler Spawning}\label{sec:ruler-spawning}
In the basic sparse ruling-set algorithm, the length of sublists between rulers may differ.
When using bulk-synchronous communication, the overall running time is therefore dominated by the length of the longest sublist.
This can be fixed using \emph{ruler spawning} (\crefrange{fig:srs-algo:line:ruler-spawning-begin}{fig:srs-algo:line:ruler-spawning-end}) introduced by Sibeyn~\cite{DBLP:conf/hipc/Sibeyn99}:
Whenever a wave reaches another ruler, we randomly choose a new ruler from the set of unreached elements and start another wave.
This ensures that there are always $r$ active waves, reducing the number of unreached vertices by $r$ in each round.
To localize the spawning process, we select the new rulers from local unreached elements and only let a wave \enquote{die} if there are none left.
Sibeyn shows that for $r \gg p\log{p}$ all elements are reached in $n/r + 1$ rounds with high probability assuming a random input~\cite{DBLP:conf/hipc/Sibeyn99}.
Hence, ruler chasing terminates after about $n/r$ rounds and is not dominated by long chains at the cost of $\approx \ln(n/r)$ times larger remaining subproblems~\cite{DBLP:conf/hipc/Sibeyn99}.
Preliminary experiments indicate that ruler spawning improves scalability, which is why we use it in our practical evaluation.

\subsection{Exploiting Input Locality}\label{sec:locality}
Many inputs emerging from applications tend to possess locality, i.e., contain sublists consisting only of local elements. %
These can be compressed to reduce the input size. In a preprocessing step, we identify \emph{local initial elements}, i.e., elements which have no non-local predecessors by a single pass over the local \succ array.
We then sequentially traverse the local lists starting from local initial elements $l$, marking all visited elements as removed, until we reach an element $j$ whose successor \succ{$j$} is non-local.
We then insert a shortcut between $l$ and \succ{$j$} by setting $\succ{$l$} \leftarrow \succ{$j$}$ and the corresponding \rank{$l$} to the sum of weights from $l$ to $\succ{j}$.
After this step, the input to the distributed algorithms consists of the local initial and terminal elements only.

To restore the updated rank and terminal elements for locally removed elements in the end, we can simply propagate the values from the local initial elements by traversing the local sublist again.
Since we only modify the successor of the local list start, we can retain all local list information by just storing the original successor for each local initial element.
The whole process of local preprocessing requires no communication, since we only access and update local elements of \succ and \rank, and runs in time $\Ohsmall{n/p}$.

While the idea of compacting local sublists has been explored for various other list-ranking algorithms before~\cite{traff_portable_1998,patel2002scalable,hameed1997contour,DBLP:journals/jea/LassousG02}, it has not been considered for sparse ruling-set with spawning before.

\subsection{Indirect Communication}\label{sec:indirection}
A challenge in scaling discrete algorithmic problems to a large number of PEs is to communicate efficiently.
Many traditional HPC applications rely on stencil communication with large amounts of data.
In contrast, communication in list ranking requires exchanging small messages between all PEs.

\begin{figure}[t]
  \centering
  \tikzset{
    compute node/.style={
      draw=my-hl1,fill=my-hl1!20
    },
    routing edge/.style={
      very thick,my-hl2
    }
  }
  \begin{subfigure}{.45\textwidth}
    \centering
    \iffigsprebuilt
      \includegraphics{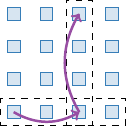}%
    \else
    \tikzsetnextfilename{fig-indirection-grid}%
    \begin{tikzpicture}
      \matrix (m) [matrix of nodes,nodes={compute node},nodes in empty cells,column sep=1em, row sep=1em] {
        &  &  & \\
        &  &  & \\
        &  &  & \\
        &  &  & \\
      };
      \node[fit=(m-4-1) (m-4-2) (m-4-3) (m-4-4),draw,dashed] {};
      \node[fit=(m-1-3) (m-2-3) (m-3-3) (m-4-3),draw,dashed] {};
      \draw[->,routing edge] (m-4-1.center) to[bend right] (m-4-3.center);
      \draw[->,routing edge] (m-4-3.center) to[bend left] (m-1-3.center);
    \end{tikzpicture}
    \fi
    \caption{Logical 2D-grid based indirection.}\label{fig:indirection:grid}
  \end{subfigure}
  \begin{subfigure}{.45\textwidth}
    \centering
    \iffigsprebuilt
      \includegraphics{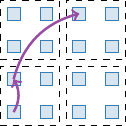}%
    \else
    \tikzsetnextfilename{fig-indirection-topo}%
    \begin{tikzpicture}
      \matrix (m) [matrix of nodes,nodes={compute node},nodes in empty cells,column sep=1em, row sep=1em] {
        &  &  & \\
        &  &  & \\
        &  &  & \\
        &  &  & \\
      };
      \node[fit=(m-1-1) (m-1-2) (m-2-1) (m-2-2),draw,dashed] {};
      \node[fit=(m-1-3) (m-1-4) (m-2-3) (m-2-4),draw,dashed] {};
      \node[fit=(m-3-1) (m-3-2) (m-4-1) (m-4-2),draw,dashed] {};
      \node[fit=(m-3-3) (m-3-4) (m-4-3) (m-4-4),draw,dashed] {};
      \draw[->,routing edge] (m-4-1.center) to[bend right] (m-3-1.center);
      \draw[->,routing edge] (m-3-1.center) to[bend left] (m-1-3.center);
    \end{tikzpicture}
    \fi
    \caption{Topology-aware indirection.}\label{fig:indirection:topology-aware}
  \end{subfigure}
  \caption{Visualization of message indirection schemes.}
\end{figure}

Sparse ruling-set only communicates a small amount of data in each round, but performs many rounds of all-to-all exchanges.
In the point-to-point model, this means that each exchange is dominated by the the startup factor $\alpha p$, which becomes increasingly impactful when scaling $p$.
To address this, we employ \emph{message indirection}: Instead of delivering all messages directly, we arrange PEs in a logical grid with side lengths of $\sqrt{p}$, assuming that $p$ is square\footnote{For arbitrary $p$, we use \texttt{MPI\_Dims\_create} to compute a balanced processor grid.}.
Then, we first forward each message to the correct processor column and then along this column to the correct row, as depicted in \cref{fig:indirection:grid}.
While this requires sending each message twice and therefore increases the communication volume by a factor of $2$, we can limit the number of communication partners per step to $\sqrt{p}$ instead of $p$, meaning that we reduce the impact of the startup factor.
The idea of indirect message delivery for reducing startups in the sparse ruling-set algorithm has been previously proposed by Sibeyn~\cite{Sibeyn1999a}, but becomes increasingly important when scaling on today's HPC systems with very fast interconnects (where $\beta$ is very small compared to $\alpha$) and increasing degree of parallelism.
It can be further generalized to $d$-dimensional grids.

\paragraph{Topology-Aware Indirection.} We propose an additional \emph{topology-aware} indirection strategy, which takes the structure of real-world systems into account.
Here we arrange PEs into $p'$ groups of size $c$, such that we have $p=cp'$.
Inside processor group $i$, we number the PEs $P_{i,0}, \ldots, P_{i,c-1}$.
To send a message from $P_{i,u}$ to $P_{j, v}$, we first send it to $P_{i,v}$, located in the same group (intra-node communication), and then to its final destination in group $j$ (inter-node communication).
We visualize this in \cref{fig:indirection:topology-aware}.
In our practical implementation, we set $c$ to the size of a single compute node, grouping PEs belonging to the same node.
This comes with two benefits:
(1) Intra-node communication has lower startup latency and transmission time.
(2) Indirection requires transmitting additional message metadata such as the destination rank for routing, that can be omitted from the final inter-node communication step.
Using fast intra-node communication when sending additional payload mitigates the increased communication volume.
Similar to grid-based based indirection, the topology-aware scheme can extended to incorporate additional dimensions, e.g, based on physical racks.

\subsection{Further Optimizations}\label{sec:further-optimizations}
\paragraph{Avoiding Communication.}\label{sec:avoid-list-inverse}
As mentioned in \cref{sec:sparse-ruling-set}, the basic sparse ruling-set algorithm ranks elements with respect to initial elements instead of terminal elements, requiring a  preprocessing step, which reverses all links in the input.
Preliminary experiments indicate that this step is costly, since each PE has to communicate all $\Ohsmall{n/p}$ elements in the worst case.
However, reversing the input can be avoided by a postprocessing step sending only $\Ohsmall{|\SetOfLists|p}$ messages in total.
At the end of the sparse ruling-set each element knows the initial element of the list. The terminal elements now send their index and rank to their initial element.
Then, all elements can request this data and adjust their final \succ and \rank values accordingly.
Since each PE only requests this information at most once per list, we can limit the number of messages to $\Ohsmall{|\SetOfLists|p}$ by locally aggregating the requests.
This can also be combined with indirection to coalesce identical requests at intermediate indirection levels.

\paragraph{Ruler Selection and Spawning.}

A new ruler can be selected by drawing random elements from the local list until an unvisited one is found.
However, when ruler-chasing is close to termination, there are much more visited than unvisited local elements, leading to many retries, each requiring a random memory access.

We avoid this by initially computing a random permutation of the local index set $\Seq{a_j}{b_j}$.
Initial ruler selection then merely requires taking the first $r/p$ elements of that permutation, skipping over terminal elements.
We keep track of the current position in the permutation. For spawning a new ruler, we continue scanning from this position until we reach an unvisited element. This allows to perform ruler selection and spawning in $\Ohsmall{n/p}$ local work.

\subsection{Scalability Analysis}\label{sec:ruler-selection}
The efficiency of sparse ruling-set directly depends on the number of rulers, since it influences both the number of communication rounds and the subproblem size.
We do not try to establish rigorous worst-case bounds on the parallel running time of the algorithm.
Rather, we make some simplifying assumptions in order to characterize the impact of algorithmic choices and locality on performance.
In particular, we would like to understand how the initial number of rulers and the degree of indirect communication should be chosen in our practical sparse ruling-set implementation.

Our main assumption is that all the performed work is well-balanced over all PEs at all time.
This can be approximated probabilistically by randomly (re-)distributing the data.
Under these assumptions, an all-to-all exchange using $d$-dimensional grid-based indirection (as outlined in \cref{sec:indirection}) with at most $h$ words per PE can be performed in time about
$\Tallall(p,h,d)=\alpha dp^{1/d}+\beta dh$.

The sparse ruling-set algorithm needs time about
\[
  T(p,n,r) =\Oh{\frac{d\beta n }{p} + \alpha dp^{1/d}\frac{n}{r} + \Tbase(p,n')}
\]
where $\Tbase$ is the running time of the base-case algorithm used, $n'$ is the remaining problem size, and $r$ denotes the number of initially selected rulers.
This equation is based on summing the cost of all-to-all communications over $n/r$ rounds.
For simplicity, we will perform a single round of sparse ruling-set and use pointer doubling as the base case, which requires parallel execution time
\[
	\Tdoubling(p,n)=\Oh{\log(n)\left(\alpha dp^{1/d}+\beta d\frac{n}{p}\right)} \text{.}
\]
With ruler spawning, the expected total number of rulers and therefore the size of the base case instance is $n'\approx r\ln(n/ r)$~\cite{DBLP:conf/hipc/Sibeyn99}.

Both $\log\frac{n}{r}$ and $\log(r\log \frac{n}{r})$ will be $\Th{\log p}$ for the range of values that will turn out to be interesting\footnote{This is wrong when $n$ is superpolynomial in $p$. However, in this case the algorithm is efficient anyway (and the input will likely not fit in memory).}
and we get
\[
  T(n,p,r) =\Oh{\frac{d\beta n }{p} + \alpha dp^{1/d}\frac{n}{r} + \alpha dp^{1/d}\log p+\beta d\frac{r\log^2 p}{p}} \text{.}
\]

Solving $\partial T/\partial r=0$, we observe the following.
\begin{observation}\label{thm:optimal-ruler-selection}
  For sparse ruling-set with pointer doubling as  base case and message indirection using $d$ levels, we achieve an optimal running time by setting $r^*=\Th{\frac{\sqrt{\alpha np^{1+1/d}/\beta}}{\log p}}$, giving us a running time of
  \[
    T^*(n, p) = T(n,p,r^*)=\Oh{\frac{d\beta n }{p} + \alpha dp^{1/d}\log p + d\sqrt{\frac{\alpha\beta n}{p^{1-1/d}}}\log p} \text{.}
  \]
\end{observation}

From that we can derive the following corollary.
\begin{corollary}
  For $\frac{n}{p}\gg \frac{\alpha}{\beta}p^{1/d}\log^2 p$ input elements per PE, the algorithm is efficient.
\end{corollary}
Thus, for smaller $n$, larger $d$ (i.e., more indirect message delivery) makes sense at the cost of larger communication overhead.

Input locality can be modeled using with a locality parameter $\delta$, denoting the ratio of elements having successor owned by the same PE\@.
If locality is perfectly balanced between PEs, applying the local preprocessing step from \cref{sec:locality} reduces the input size and total running time to $T^*((1-\delta)n,p) + \Ohsmall{n/p}$.
Unbalanced locality can be handled by randomly redistributing the data after local preprocessing.
However, this only saves a constant factor on running time, since the bottleneck communication volume is still $\Ohsmall{n/p}$ in the worst case.

\section{Experimental Evaluation}\label{sec:evaluation}
\paragraph{Experimental Setup.}\label{sec:experimental-setup}
We implement our algorithm variants using C++ and provide an open-source implementation\footnote{\url{https://github.com/niklas-uhl/kascade}}.
We run our experiments on the thin nodes of SuperMUC-NG. %
Each node is equipped with an Intel Skylake Xeon Platinum 8174 processor with 48 cores.
The available memory per node is limited to 96 GB\@.
The interconnect is an OmniPath network with 100 Gbit/s.
Our code is built on top of the C++ MPI interface KaMPIng~\cite{Uhl2024}, is compiled using g++-15.1.0 and Intel MPI 2021.15.0, and uses optimization flags \texttt{-O3} and \texttt{-march=native}.

\paragraph{Methodology.}\label{sec:methodology}
For each experiment, we generate input instances and perform 5 iterations, where we discard the first run since it might incur communication setup overhead.
We report the mean total running time (excluding input generation) over the remaining iterations.
Errors bars indicate the minimum and maximum running time observed on a specific configuration.
We perform \emph{weak-scaling} experiments, where we fix the input size $n/p$ per MPI process and scale it proportional to the number of MPI processes $p$.

We choose the number of rulers based on the results from our analysis, scaling the local number of rulers per PE relative to the local input size per PE.
Using a parameter study, we determined suitable constant factors. %
We run two rounds of sparse ruling-set and use pointer doubling as the base case algorithm, which proved to be more efficient than a single round.
Additional experiments indicate that more than two rounds do not pay off.
We engineered the pointer-doubling implementation to apply the same optimizations as our sparse ruling-set algorithm, with support for message indirection and locality exploitation.

\paragraph{Input Instances.}\label{sec:input-instances}
We use two classes of input instances.
First, we generate lists of size $n$, where we set $\succ{$i$} = i + 1$.
Then the parameter $\gamma \in [0, 1]$ controls the degree of locality: We sample $\gamma n$ indices from $\Seq{0}{n-1}$, which we randomly permute.
This means that for $\gamma=0.0$ we get a list where each PE is assigned a contiguous sub-list, while $\gamma=1.0$ fully permutes the list such that there is no locality.
We denote these instances as $\textsc{List}(n/p, \gamma)$.

We derive additional instances from random graphs.
We use the communication free graph generators from KaGen~\cite{DBLP:journals/jpdc/FunkeLMPSSSL19} to generate random Erdős-Rényi graphs (\textsc{GNM}) and two-dimensional random geometric graphs (\textsc{RGG2D}).
We then perform a breadth-first search to construct a tree, which we transform to a Euler-tour~\cite{Jaja1992}.
This allows us to generate lists which follow the locality properties of the input graphs.
\subsection{Exploiting Input Locality}\label{sec:exp-locality}

\begin{figure}[t]
  \centering
  \iffigsprebuilt
    \includegraphics{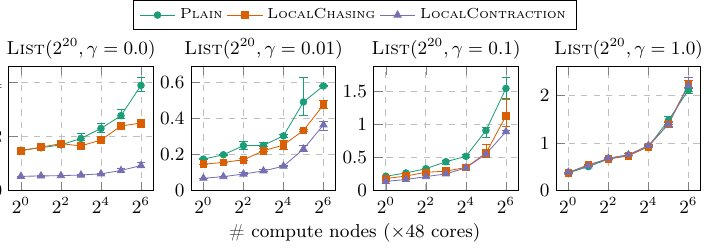}%
  \else
  \tikzsetnextfilename{fig-locality}%
  \begin{tikzpicture}
    \begin{groupplot}[
      kascade plot format,
      group style={
        columns=4,
        xlabels at=edge bottom,
        ylabels at=edge left,
        horizontal sep=2em,
      },
      width=.2\linewidth,
      scale only axis,
      xticklabel={$2^{\pgfmathprintnumber[fixed,precision=0]{\tick}}$},
      ylabel={time $/s$},
      ymin=0,
      xmajorgrids=true,
      ymajorgrids=true,
      grid style=dashed,
      legend cell align=left,
      legend columns=-1,
      legend style={
        font=\scriptsize
      },
      label style={font=\footnotesize},
      ytick pos=left,
      y label style={overlay},
      y tick label style={overlay},
      tick label style={font=\footnotesize},
      every axis plot/.style={
        error bars/y dir=both,
        error bars/y explicit,
      },
      every axis title/.append style={
        at={(0.5, 0.90)}
      }
      ]
      \nextgroupplot[title={$\textsc{List}(2^{20},\gamma=0.0)$},legend to name=locality legend]
      \datafileKV{prefix=plots/locality/locality,graph=1,config=4}
      \addplot+ table {\kascadedatafilename};
      \addlegendentry{\textsc{Plain}}
      \datafileKV{prefix=plots/locality/locality,graph=1,config=3}
      \addplot+ table {\kascadedatafilename};
      \addlegendentry{\textsc{LocalChasing}}
      \datafileKV{prefix=plots/locality/locality,graph=1,config=2}
      \addplot+ table {\kascadedatafilename};
      \addlegendentry{\textsc{LocalContraction}}
      
      \nextgroupplot[title={$\textsc{List}(2^{20},\gamma=0.01)$}]
      \datafileKV{prefix=plots/locality/locality,graph=3,config=4}
      \addplot+ table {\kascadedatafilename};
      \datafileKV{prefix=plots/locality/locality,graph=3,config=3}
      \addplot+ table {\kascadedatafilename};
      \datafileKV{prefix=plots/locality/locality,graph=3,config=2}
      \addplot+ table {\kascadedatafilename};

      \nextgroupplot[title={$\textsc{List}(2^{20},\gamma=0.1)$}]
      \datafileKV{prefix=plots/locality/locality,graph=4,config=4}
      \addplot+ table {\kascadedatafilename};
      \datafileKV{prefix=plots/locality/locality,graph=4,config=3}
      \addplot+ table {\kascadedatafilename};
      \datafileKV{prefix=plots/locality/locality,graph=4,config=2}
      \addplot+ table {\kascadedatafilename};

      \nextgroupplot[title={$\textsc{List}(2^{20},\gamma=1.0)$}]
      \datafileKV{prefix=plots/locality/locality,graph=2,config=4}
      \addplot+ table {\kascadedatafilename};
      \datafileKV{prefix=plots/locality/locality,graph=2,config=3}
      \addplot+ table {\kascadedatafilename};
      \datafileKV{prefix=plots/locality/locality,graph=2,config=2}
      \addplot+ table {\kascadedatafilename};
    \end{groupplot}
    \node[fit=(group c1r1) (group c4r1)] (bb) {};
    \node[below=1em of bb,font=\footnotesize] {\# compute nodes ($\times 48$ cores)};
    \node[inner sep=0,above=1.5em of bb]  {\pgfplotslegendfromname{locality legend}};
  \end{tikzpicture}
  \fi
  \caption{Evaluation of different locality-aware techniques on random lists with varying degree of locality. We do not use message indirection.}
  \label{fig:exp-locality}
\end{figure}

In \cref{fig:exp-locality}, we investigate our different locality-aware optimizations for the sparse ruling-set algorithm.
The \textsc{Plain} variant does not employ any optimizations for local elements.
The \textsc{LocalityChasing} variant  provides a straightforward way to exploit locality: it runs the distributed algorithm but follows local ruler chains until reaching a non-local successor before participating in the next communication round.
The \textsc{LocalContraction} variant performs the local sublist contraction step described in \cref{sec:locality}, before running the actual distributed sparse ruling-set algorithm.
This has the advantage that we can use the effective number of elements as a parameter for ruler selection. %
We see that these two optimization techniques are highly effective for inputs with a high degree of locality.
For $\gamma=0.0$, \textsc{LocalContraction} is up to $4\times$ faster than \textsc{Plain}.
Furthermore, we see that these optimizations add almost no overhead in case there is no locality in the input ($\gamma=1.0$).
Therefore, we enable \textsc{LocalContraction} for all following experiments.

\subsection{Scaling Experiments}\label{sec:exp-indirection}
\begin{figure}[t]
  \centering
  \iffigsprebuilt
    \includegraphics{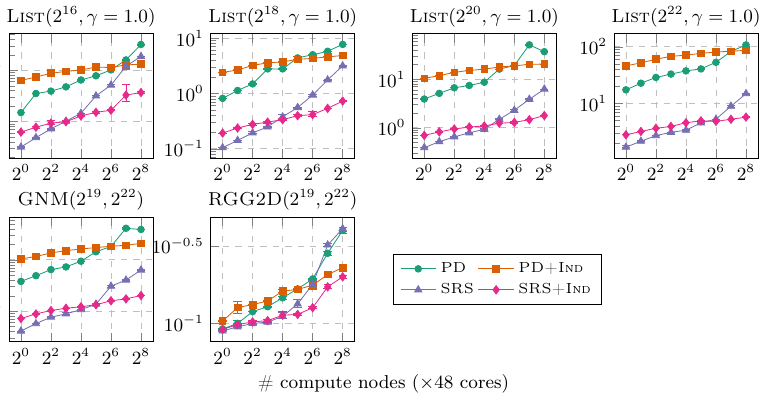}%
  \else
  \tikzsetnextfilename{fig-scalability}%
  \begin{tikzpicture}
    \begin{groupplot}[
      kascade plot format,
      group style={
        columns=4,
        rows=2,
        xlabels at=edge bottom,
        ylabels at=edge left,
        horizontal sep=3em,
      },
      width=.2\linewidth,
      scale only axis,
      xticklabel={$2^{\pgfmathprintnumber[fixed,precision=0]{\tick}}$},
      ylabel={time $/s$},
      ymode=log,
      xmajorgrids=true,
      ymajorgrids=true,
      grid style=dashed,
      legend cell align=left,
      legend columns=2,
      legend style={
        font=\scriptsize
      },
      label style={font=\footnotesize},
      ytick pos=left,
      y label style={overlay},
      y tick label style={overlay},
      tick label style={font=\footnotesize},
      every axis plot/.style={
        error bars/y dir=both,
        error bars/y explicit,
      },
      every axis title/.append style={
        at={(0.5, 0.90)}
      }
      ]
      \nextgroupplot[title={$\textsc{List}(2^{16}, \gamma=1.0)$},legend to name=scalability legend]
      \datafileKV{graph=1,config=1,prefix=plots/scalability/permuted_list/scalability-permuted-list}
      \addplot+ table {\kascadedatafilename};
      \addlegendentry{\textsc{PD}}
      \datafileKV{graph=1,config=2,prefix=plots/scalability/permuted_list/scalability-permuted-list}
      \addplot+ table {\kascadedatafilename};
      \addlegendentry{\textsc{PD+Ind}}
      \datafileKV{graph=1,config=4,prefix=plots/scalability/permuted_list/scalability-permuted-list}
      \addplot+ table {\kascadedatafilename};
      \addlegendentry{\textsc{SRS}}
      \datafileKV{graph=1,config=3,prefix=plots/scalability/permuted_list/scalability-permuted-list}
      \addplot+ table {\kascadedatafilename};
      \addlegendentry{\textsc{SRS+Ind}}

      \nextgroupplot[title={$\textsc{List}(2^{18}, \gamma=1.0)$}]
      \datafileKV{graph=4,config=1,prefix=plots/scalability/permuted_list/scalability-permuted-list}
      \addplot+ table {\kascadedatafilename};
      \datafileKV{graph=4,config=2,prefix=plots/scalability/permuted_list/scalability-permuted-list}
      \addplot+ table {\kascadedatafilename};
      \datafileKV{graph=4,config=4,prefix=plots/scalability/permuted_list/scalability-permuted-list}
      \addplot+ table {\kascadedatafilename};
      \datafileKV{graph=4,config=3,prefix=plots/scalability/permuted_list/scalability-permuted-list}
      \addplot+ table {\kascadedatafilename};

      \nextgroupplot[title={$\textsc{List}(2^{20}, \gamma=1.0)$}]
      \datafileKV{graph=8,config=1,prefix=plots/scalability/permuted_list/scalability-permuted-list}
      \addplot+ table {\kascadedatafilename};
      \datafileKV{graph=8,config=2,prefix=plots/scalability/permuted_list/scalability-permuted-list}
      \addplot+ table {\kascadedatafilename};
      \datafileKV{graph=8,config=4,prefix=plots/scalability/permuted_list/scalability-permuted-list}
      \addplot+ table {\kascadedatafilename};
      \datafileKV{graph=8,config=3,prefix=plots/scalability/permuted_list/scalability-permuted-list}
      \addplot+ table {\kascadedatafilename};

      \nextgroupplot[title={$\textsc{List}(2^{22}, \gamma=1.0)$}]
      \datafileKV{graph=7,config=1,prefix=plots/scalability/permuted_list/scalability-permuted-list}
      \addplot+ table {\kascadedatafilename};
      \datafileKV{graph=7,config=2,prefix=plots/scalability/permuted_list/scalability-permuted-list}
      \addplot+ table {\kascadedatafilename};
      \datafileKV{graph=7,config=4,prefix=plots/scalability/permuted_list/scalability-permuted-list}
      \addplot+ table {\kascadedatafilename};
      \datafileKV{graph=7,config=3,prefix=plots/scalability/permuted_list/scalability-permuted-list}
      \addplot+ table {\kascadedatafilename};
      
      \nextgroupplot[title={$\textsc{GNM}(2^{19},2^{22})$}]
      \datafileKV{graph=2,config=1,prefix=plots/scalability/euler/scalability-euler}
      \addplot+ table {\kascadedatafilename};
      \datafileKV{graph=2,config=2,prefix=plots/scalability/euler/scalability-euler}
      \addplot+ table {\kascadedatafilename};
      \datafileKV{graph=2,config=4,prefix=plots/scalability/euler/scalability-euler}
      \addplot+ table {\kascadedatafilename};
      \datafileKV{graph=2,config=3,prefix=plots/scalability/euler/scalability-euler}
      \addplot+ table {\kascadedatafilename};

      \nextgroupplot[title={$\textsc{RGG2D}(2^{19}, 2^{22})$}]
      \datafileKV{graph=3,config=1,prefix=plots/scalability/euler/scalability-euler}
      \addplot+ table {\kascadedatafilename};
      \datafileKV{graph=3,config=2,prefix=plots/scalability/euler/scalability-euler}
      \addplot+ table {\kascadedatafilename};
      \datafileKV{graph=3,config=4,prefix=plots/scalability/euler/scalability-euler}
      \addplot+ table {\kascadedatafilename};
      \datafileKV{graph=3,config=3,prefix=plots/scalability/euler/scalability-euler}
      \addplot+ table {\kascadedatafilename};

      \nextgroupplot[group/empty plot]
      \nextgroupplot[group/empty plot]
      
    \end{groupplot}
    \node[fit=(group c1r2) (group c4r2)] (bb) {};
    \node[below=1em of bb,font=\footnotesize] {\# compute nodes ($\times 48$ cores)};
    \node[inner sep=0,right=2em of group c2r2]  {\pgfplotslegendfromname{scalability legend}};
  \end{tikzpicture}
  \fi
  \caption{Scalability of pointer doubling and sparse ruling-set on different instances.}
  \label{fig:exp-scalability}
\end{figure}

\cref{fig:exp-scalability} shows the running times of the sparse ruling-set (\textsc{SRS}) and the pointer doubling (\textsc{PD}) algorithms on random list instances with $n/p \in \{2^{16}, 2^{18}, 2^{20} 2^{22}\}$ elements and on two Euler tour instances with $\approx n/p = 2^{20}$ elements, constructed from graphs with $n/p=2^{19}$ vertices and $m/p=2^{22}$ edges.
We run each algorithm in two variants -- one using direct communication and one using topology-aware indirect communication enabled (\textsc{+Ind} suffix, see \cref{sec:indirection} for details).

Overall, we see that \textsc{SRS+Ind} scales well to even the largest configurations ($12\,228$ cores), especially for $n/p \in \{2^{20}, 2^{22}\}$.
The sparse ruling-set variants are up to $15\times$ faster than pointer doubling. %
As expected, indirect communication pays off for both algorithms as the number of cores increases.
For $n/p=2^{16}$ elements, \textsc{SRS+Ind} becomes faster from $768$ cores on.
For larger input sizes the break-even shifts to higher core counts, as the larger communication volume hides the latency costs.
Nevertheless, even for $n/p=2^{22}$ elements \textsc{SRS+Ind} outperforms \textsc{SRS} from $3\,072$ cores on.

For pointer doubling, direct communication remains preferable up to higher core counts than for sparse ruling-set, as the larger amount of work performed by the algorithm, combined with a smaller number of communication rounds, hides the latency costs for longer.
With further increases in the core count, indirect communication eventually becomes faster as well.

Across all instances and core counts, the better variant of sparse ruling-set is always faster than the better pointer-doubling variant for the respective configuration.
On the \textsc{RGG2D}-based instances, \textsc{PD+Ind} is competitive with \textsc{SRS+Ind}.
These inputs have high locality, which only leaves about one percent of the initial problem after local preprocessing. %
Here, both \textsc{PD+Ind} and \textsc{SRS+Ind} are dominated by latency.
\textsc{GNM}-based instances show results similar to random lists, since the graphs possess almost no locality.

Additionally, we scaled \textsc{SRS+Ind} to $24\,576$ cores on a reduced set of inputs.
In this configuration, \textsc{SRS+Ind} requires only 2.5 and 6.7 seconds on random lists with $n/p = 2^{20}$ and $n/p = 2^{22}$ elements, respectively, ranking over 100 billion elements in the latter case. %

\begin{figure}[t]
  \centering
  \iffigsprebuilt
    \includegraphics{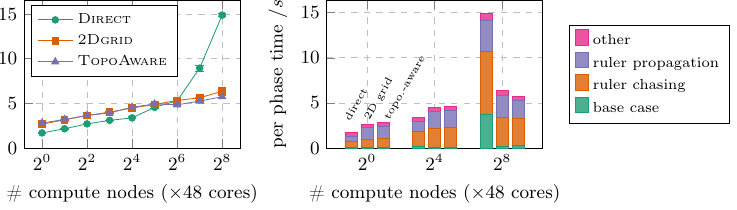}%
  \else
  \tikzsetnextfilename{fig-indirection-impact}%
  \begin{tikzpicture}[
    base case/.style={
      fill=Dark2-A!80,
      draw=Dark2-A,
    },
    ruler chasing/.style={
      fill=Dark2-B!80,
      draw=Dark2-B,
    },
    ruler propagation/.style={
      fill=Dark2-C!80,
      draw=Dark2-C,
    },
    rest/.style={
      fill=Dark2-D!80,
      draw=Dark2-D,
    },
    ]
    \begin{groupplot}[
      group style={
        columns=3,
        xlabels at=edge bottom,
        ylabels at=edge left,
        horizontal sep=4.5em,
      },
      width=.3\linewidth,
      height=2.5cm,
      scale only axis,
      xticklabel={$2^{\pgfmathprintnumber[fixed,precision=0]{\tick}}$},
      xlabel={\# compute nodes ($\times 48$ cores)},
      ylabel={time $/s$},
      ymin=0,
      xmajorgrids=true,
      ymajorgrids=true,
      grid style=dashed,
      legend cell align=left,
      legend style={
        font=\scriptsize
      },
      label style={font=\footnotesize},
      ytick pos=left,
      y label style={overlay},
      y tick label style={overlay},
      tick label style={font=\footnotesize},
      ]

      \nextgroupplot[kascade plot format,
      error bars/y dir=both,
      error bars/y explicit,
      legend pos=north west,
      ]
      \datafileKV{graph=4,config=3,prefix=plots/indirection/indirection}
      \addplot+ table {\kascadedatafilename};
      \addlegendentry{\textsc{Direct}}
      \datafileKV{graph=4,config=1,prefix=plots/indirection/indirection}
      \addplot+ table {\kascadedatafilename};
      \addlegendentry{\textsc{2Dgrid}}
      \datafileKV{graph=4,config=2,prefix=plots/indirection/indirection}
      \addplot+ table {\kascadedatafilename};
      \addlegendentry{\textsc{TopoAware}}

      \nextgroupplot[ybar stacked,
      xtick=data,
      xticklabels={$2^0$, $2^4$, $2^8$},
      ylabel={per phase time $/ s$},
      enlarge x limits=0.3,
      bar width=1.5ex,
      reverse legend,
      legend to name=phase legend,
      ]
      \addplot[bar shift=-2ex,base case] table [x expr=\coordindex,y=base_case,col sep=comma] {plots/barplot/bar_c0.csv};
      \addlegendentry{base case}
      \addplot[bar shift=-2ex,ruler chasing] table [x expr=\coordindex,y=chase_rulers,col sep=comma] {plots/barplot/bar_c0.csv};
      \addlegendentry{ruler chasing}
      \addplot[bar shift=-2ex,ruler propagation] table [x expr=\coordindex,y=ruler_propagation,col sep=comma] {plots/barplot/bar_c0.csv};
      \addlegendentry{ruler propagation}
      \addplot[bar shift=-2ex,rest] table [x expr=\coordindex,y=rest,col sep=comma] {plots/barplot/bar_c0.csv};
      \addlegendentry{other}
      
      \makeatletter
      \pgfplots@stacked@isfirstplottrue
      \makeatother
      
      \addplot[base case] table [x expr=\coordindex,y=base_case,col sep=comma] {plots/barplot/bar_c1.csv};
      \addplot[ruler chasing] table [x expr=\coordindex,y=chase_rulers,col sep=comma] {plots/barplot/bar_c1.csv};
      \addplot[ruler propagation] table [x expr=\coordindex,y=ruler_propagation,col sep=comma] {plots/barplot/bar_c1.csv};
      \addplot[rest] table [x expr=\coordindex,y=rest,col sep=comma] {plots/barplot/bar_c1.csv};

      \makeatletter
      \pgfplots@stacked@isfirstplottrue
      \makeatother
      
      \addplot[bar shift=2ex,base case] table [x expr=\coordindex,y=base_case,col sep=comma] {plots/barplot/bar_c2.csv};
      \addplot[bar shift=2ex,ruler chasing] table [x expr=\coordindex,y=chase_rulers,col sep=comma] {plots/barplot/bar_c2.csv};
      \addplot[bar shift=2ex,ruler propagation] table [x expr=\coordindex,y=ruler_propagation,col sep=comma] {plots/barplot/bar_c2.csv};
      \addplot[bar shift=2ex,rest] table [x expr=\coordindex,y=rest,col sep=comma] {plots/barplot/bar_c2.csv};

      \node[font=\tiny,rotate=60,anchor=west] at (axis cs:-0.35,2.5) {direct};
      \node[font=\tiny,rotate=60,anchor=west] at (axis cs:-0.05,2.5) {2D grid};
      \node[font=\tiny,rotate=60,anchor=west] at (axis cs:0.25,2.5) {topo.-aware};

      \nextgroupplot[group/empty plot]
    \end{groupplot}
    \node[right=1em of group c2r1] {\pgfplotslegendfromname{phase legend}};
  \end{tikzpicture}
  \fi
\caption{Impact of different message indirection techniques on $\textsc{List}(2^{22}, \gamma = 1.0)$ and detailed phase breakdown.}
  \label{fig:exp-indirection}
\end{figure}

\paragraph{Performance Breakdown.}
In \cref{fig:exp-indirection}, we further analyze the impact of the different indirection techniques and give a more detailed insight into the running time of the main three phases of the sparse ruling-set algorithm.
We show absolute running times for sparse ruling-set (with and without indirect communication) on random ($\gamma=1.0$) lists with $n/p=2^{22}$ elements (left) and break them down into the ruler-chasing, ruler-propagation and base-case phases (right).
We compare the 2D-grid and topology-aware indirection strategies introduced in \cref{sec:indirection}.

We see that topology-aware communicator grouping has a positive but limited effect on the running time, by up to $10$ percent.
Further analysis explains why this effect is not more pronounced:
While intra-node communication is substantially (up to $1.5\times$ -- $3\times$) faster than communicating between distant compute nodes, the 2D-grid based indirection scheme still groups physically close processors, where communication performance is almost as fast as inside a node.

The plot on the right shows that, for small core configurations, ruler chasing and ruler propagation dominate the running time.
With increasing core count, the time spent in the base case increases.
This is expected as the size of the base-case subproblem, i.e., the number of selected rulers, grows with the total problem size while the local input size remains constant in this weak-scaling experiment.

\section{Conclusion and Future Work}\label{sec:conclusion}
We present a carefully engineered distributed-memory implementation of the sparse ruling-set algorithm and evaluate it on up to 24\,576 processors across a variety of instances with differing characteristics, increasing the scale of previous experimental studies by two orders of magnitude.
Furthermore, we provide detailed insights on the effects of the applied optimization techniques, which are of independent interest for designing scalable algorithms for other discrete problems.

We also conducted preliminary experiments with low-latency asynchronous communication and MPI's remote-direct memory access (RMA).
While the asynchronous approach showed promising results when the number of communication partners is low, and can hide latency by avoiding hard synchronization, it requires further engineering to match the performance of our highly tuned synchronous variants.
Limited support for highly irregular accesses by MPI's RMA interface prohibits straightforward implementations.

In the future, we want to adapt our techniques to the more general tree rooting problem, which has applications in e.g. distributed MST computations.
While indirect communication and locality exploitation also apply here, tree rooting requires revised ruler selection and needs to consider contention for elements with many predecessors.
Lastly, scaling to higher processor counts will require additional levels of indirection to effectively tame startup latencies.

\begin{credits}
  \subsubsection{Acknowledgments and Artifact Availability.}
  The authors gratefully acknowledge the Gauss Centre for Supercomputing e.V. (\url{www.gauss-centre.eu}) for funding this project by providing computing time on the GCS Supercomputer SuperMUC-NG at Leibniz Supercomputing Centre (\url{www.lrz.de}).
  The authors gratefully acknowledge the computing time provided on the high-performance computer HoreKa by the National High-Performance Computing Center at KIT (NHR@KIT). This center is jointly supported by the Federal Ministry of Education and Research and the Ministry of Science, Research and the Arts of Baden-Württemberg, as part of the National High-Performance Computing (NHR) joint funding program (\url{https://www.nhr-verein.de/en/our-partners}). HoreKa is partly funded by the German Research Foundation (DFG).
  The artifact is available in the Zenodo repository~\cite{reproducibility}.
\end{credits}

\bibliographystyle{splncs04}
\bibliography{kascade.bib}

\begin{thebibliography}{10}
\providecommand{\url}[1]{\texttt{#1}}
\providecommand{\urlprefix}{URL }
\providecommand{\doi}[1]{https://doi.org/#1}

\bibitem{DBLP:journals/ipl/AndersonM90}
Anderson, R.J., Miller, G.L.: A simple randomized parallel algorithm for
  list-ranking. Inf. Process. Lett.  \textbf{33}(5),  269--273 (1990).
  \doi{10.1016/0020-0190(90)90196-5}

\bibitem{anderson1991deterministic}
Anderson, R.J., Miller, G.L.: Deterministic parallel list ranking. Algorithmica
   \textbf{6}(6),  859--868 (1991). \doi{10.1007/BF01759076}

\bibitem{DBLP:conf/focs/ColeV86}
Cole, R., Vishkin, U.: Approximate and exact parallel scheduling with
  applications to list, tree and graph problems. In: {FOCS}. pp. 478--491.
  {IEEE} Computer Society (1986). \doi{10.1109/SFCS.1986.10}

\bibitem{dehne1997randomized}
Dehne, F., Song, S.W.: Randomized parallel list ranking for distributed memory
  multiprocessors. Int. J. Parallel Program.  \textbf{25}(1),  1--16 (1997).
  \doi{10.1007/BF02700044}

\bibitem{dehne2002efficient}
Dehne, F.K.H.A., et~al.: Efficient parallel graph algorithms for coarse-grained
  multicomputers and {BSP}. Algorithmica  \textbf{33}(2),  183--200 (2002).
  \doi{10.1007/S00453-001-0109-4}

\bibitem{DBLP:journals/jpdc/FunkeLMPSSSL19}
Funke, D., et~al.: Communication-free massively distributed graph generation.
  J. Parallel Distributed Comput.  \textbf{131},  200--217 (2019).
  \doi{10.1016/J.JPDC.2019.03.011}

\bibitem{hameed1997contour}
Hameed, F., et~al.: Contour ranking on coarse grained machines: a case study
  for low-level vision computations. Concurr. Pract. Exp.  \textbf{9}(3),
  203--221 (1997)

\bibitem{Jaja1992}
J{\'{a}}J{\'{a}}, J.F.: An Introduction to Parallel Algorithms. Addison-Wesley
  (1992)

\bibitem{DBLP:journals/jea/LassousG02}
Lassous, I.G., Gustedt, J.: Portable list ranking: An experimental study. {ACM}
  J. Exp. Algorithmics  \textbf{7}, ~7 (2002). \doi{10.1145/944618.944625}

\bibitem{miller1985parallel}
Miller, G.L., Reif, J.H.: Parallel tree contraction and its application. In:
  {FOCS}. pp. 478--489. {IEEE} Computer Society (1985).
  \doi{10.1109/SFCS.1985.43}

\bibitem{Pan2020}
Pan, T., et~al.: Fast de bruijn graph compaction in distributed memory
  environments. {IEEE} {ACM} Trans. Comput. Biol. Bioinform.  \textbf{17}(1),
  136--148 (2020). \doi{10.1109/TCBB.2018.2858797}

\bibitem{patel2002scalable}
Patel, J.N., et~al.: Scalable parallel implementations of list ranking on
  fine-grained machines. {IEEE} Trans. Parallel Distributed Syst.
  \textbf{8}(10),  1006--1018 (1997). \doi{10.1109/71.629484}

\bibitem{fast_and_scalable_gpu}
Rehman, M.S., et~al.: Fast and scalable list ranking on the {GPU}. In: {ICS}.
  pp. 235--243. {ACM} (2009). \doi{10.1145/1542275.1542311}

\bibitem{reid1994list}
Reid{-}Miller, M.: List ranking and list scan on the cray {C-90}. In: {SPAA}.
  pp. 104--113. {ACM} (1994). \doi{10.1145/181014.181049}

\bibitem{reproducibility}
Sanders, P., Schimek, M., Uhl, T.N., Weidmann, T.: Artifact of the paper:
  Engineering scalable distributed list ranking (2026).
  \doi{10.5281/zenodo.20509834}

\bibitem{SMDD19}
Sanders, P., et~al.: Sequential and Parallel Algorithms and Data Structures -
  The Basic Toolbox. Springer (2019). \doi{10.1007/978-3-030-25209-0}

\bibitem{sibeyn_better_1997}
Sibeyn, J.F.: Better trade-offs for parallel list ranking. In: {SPAA}. pp.
  221--230. {ACM} (1997). \doi{10.1145/258492.258514}

\bibitem{DBLP:conf/hipc/Sibeyn99}
Sibeyn, J.F.: Ultimate parallel list ranking? In: HiPC. Lecture Notes in
  Computer Science, vol.~1745, pp. 197--201. Springer (1999).
  \doi{10.1007/978-3-540-46642-0_28}

\bibitem{sibeyn_list-ranking_2003}
Sibeyn, J.F.: List-ranking on interconnection networks. Inf. Comput.
  \textbf{181}(2),  75--87 (2003). \doi{10.1016/S0890-5401(02)00029-9}

\bibitem{sibeyn2003minimizing}
Sibeyn, J.F.: Minimizing global communication in parallel list ranking. In:
  Euro-Par. Lecture Notes in Computer Science, vol.~2790, pp. 894--902.
  Springer (2003). \doi{10.1007/978-3-540-45209-6_123}

\bibitem{Sibeyn1999a}
Sibeyn, J.F., Guillaume, F., Seidel, T.: Practical parallel list ranking. J.
  Parallel Distributed Comput.  \textbf{56}(2),  156--180 (1999).
  \doi{10.1006/JPDC.1998.1508}

\bibitem{traff_portable_1998}
Tr{\"{a}}ff, J.L.: Portable randomized list ranking on multiprocessors using
  {MPI}. In: {PVM/MPI}. Lecture Notes in Computer Science, vol.~1497, pp.
  395--402. Springer (1998). \doi{10.1007/BFB0056600}

\bibitem{Uhl2024}
Uhl, T.N., et~al.: Kamping: Flexible and (near) zero-overhead {C++} bindings
  for {MPI}. In: {SC}. p.~44. {IEEE} (2024). \doi{10.1109/SC41406.2024.00050}

\bibitem{wei2012optimization}
Wei, Z., J{\'{a}}J{\'{a}}, J.F.: Optimization of linked list prefix
  computations on multithreaded gpus using {CUDA}. Parallel Process. Lett.
  \textbf{22}(4) (2012). \doi{10.1142/S0129626412500120}

\bibitem{wyllie1979complexity}
Wyllie, J.: The Complexity of Parallel Computations. Ph.D. thesis, Cornell
  University, {USA} (1979)

\end{thebibliography}

\end{document}